\documentclass[aps,prl,
   amsfonts,amssymb,amsmath,
   twocolumn,
   superscriptaddress,
   noshowpacs,
   preprintnumbers,
   amsmath,amssymb,
]{revtex4-1}

\usepackage{color}
\usepackage{graphicx}
\usepackage[pdftex,colorlinks=true,linkcolor=red,citecolor=blue]{hyperref}

\begin{document}

\title{Local inhomogeneity and filamentary superconductivity in Pr-doped CaFe$_{2}$As$_{2}$}

\author{Krzysztof Gofryk}
\author{Minghu~Pan}
\author{Claudia~Cantoni}
\author{Bayrammurad~Saparov}
\author{Jonathan~E.~Mitchell}
\author{Athena~S.~Sefat}
\affiliation{Oak Ridge National Laboratory, Oak Ridge, Tennessee 37831, USA}

\date{\today}

\begin{abstract}

We use multi-scale techniques to determine the extent of local inhomogeneity and superconductivity in Ca$_{0.86}$Pr$_{0.14}$Fe$_{2}$As$_{2}$ single crystal. The inhomogeneity is manifested as a spatial variation of praseodymium concentration, local density of states, and superconducting order parameter. We show that the high-$T_{c}$ superconductivity emerges from clover-like defects associated with Pr dopants. The highest $T_{c}$ is observed in both the tetragonal and collapsed tetragonal phases, and its filamentary nature is a consequence of non-uniform Pr distribution that develops localized, isolated superconducting regions within the crystals.

\end{abstract}

\maketitle


\emph{Introduction.} -- The discovery of unconventional superconductivity in Fe-based superconductors \cite{Kamihara,chen} opened a new era in superconductivity research \cite{greg}. These new superconductors share several characteristics with their high-$T_{c}$ relatives based on copper (cuprates) \cite{zl}. First, both are \textit{d}-electron materials having layered structures with characteristic CuO$_{2}$ and FeAs/FeSe planes where the superconductivity is believed to originate. Second, their parent compounds are antiferromagnetic and superconductivity arises by application of pressure or when they are doped by electrons or holes. Aside from introducing the charge carriers, the role of doping in iron-based superconductors, as well as in cuprates and heavy fermions, is still unclear, nor it is well understood how the dopants are distributed in the material. In addition, to promote superconductivity, the dopants are potential sources of inhomogeneity such as phase separation and crystalline or electronic disorder. In many models of superconductivity, it is assumed that the dopant atoms are distributed uniformly in the material. However, there are many indications for nanoscale inhomogeneity in these materials \cite{a,b}.

The issues of doping and inhomogeneity are well exemplified in Ca$_{1-x}$Pr$_{x}$Fe$_{2}$As$_{2}$ superconductor \cite{lv,saha}. This electron doped system can show $T_{c}$~$\approx$~45~K, the highest among the pnictides with ThCr$_{2}$Si$_{2}$ crystal structure \cite{122a,122b,122c}. Such a high $T_{c}$ in a simple structure stimulated a huge scientific interest in this material from both, fundamental and applied point of view. Despite extensive studies, the origin of the high-$T_{c}$ superconducting state in Pr-doped CaFe$_{2}$As$_{2}$ and its filamentary nature are still under debate (see Refs.~\onlinecite{qi,lv2,ma,6,wei}).

Here we address these problems by investigating the local electronic inhomogeneous state in Pr-doped CaFe$_{2}$As$_{2}$. We perform extensive electronic and structural studies of single crystals of Ca$_{0.86}$Pr$_{0.14}$Fe$_{2}$As$_{2}$ superconductor (onset $T_{c}$~=~45~K) by use of macro- (magnetic susceptibility, electrical resistivity, specific heat) and micro-scale (scanning transmission electron microscopy [STEM] coupled with electron energy loss spectroscopy [EELS] and scanning tunneling microscopy [STM]) measurements. EELS results indicate that the Pr distribution is not uniform, while STM shows an electronic inhomogeneity, which is evidenced as a spatial variation of both local density of states and the superconducting order parameter. The results suggest that the highest $T_{c}$, associated with clover-like defect, resides in the close vicinity of the Pr atoms and forms isolated superconducting regions. STM also confirms that a significant part of the sample (up to 30~\%) remains in the normal state when cooled below $T_{c}$, in agreement with bulk studies. Furthermore, we find that the high-$T_{c}$ superconductivity in Ca$_{0.86}$Pr$_{0.14}$Fe$_{2}$As$_{2}$ is observed in both tetragonal and collapsed tetragonal phases. We discuss the implications of this study on the role of disorder and its relationship with  superconductivity, and give perspectives with extension to other exotic superconductors.

\emph{Methods.} -- Single crystals of Ca$_{0.86}$Pr$_{0.14}$Fe$_{2}$As$_{2}$ were grown out of FeAs flux with the typical size of about 2$\times$1.5$\times$0.2 mm$^{3}$ \cite{122b}. The energy-dispersive spectroscopy (EDS) analysis points to inhomogeneous distribution of Pr dopant on the micron size. Several spots on different crystals from the same batch have been used for the analysis giving a heterogenous Pr concentration ranging from 11.75 to 15.33 \%. The average concentration has been established as 14 \%, which is used in the paper. The magnetic susceptibility were measured using a Quantum Design MPMS-7 device. The electrical resistivity and heat capacity were measured using a four wire and relaxation methods, respectively, implemented in a Quantum Design PPMS-14 setup. The low temperature diffraction experiments were carried out using a PANalytical X'Pert PRO MPD x-ray diffractometer with Cu-K$_{\alpha 1}$ radiation. STEM and EELS were performed using a Nion UltraSTEM~200 microscope. The samples were imaged using the high angle annular dark field detector (HAADF), which selects diffracted electrons that have undergone elastic scattering in close proximity to the nuclei yielding an intensity nearly proportional to $Z^{2}$ and consequent chemical information. The STM and scanning tunneling spectroscopy (STS) experiments were carried out in a home-built low temperature scanning tunneling microscope system. The samples were cleaved in ultra-high vacuum and then loaded into the STM head for investigations at low temperatures.

\begin{figure}[b!]
\begin{centering}
\includegraphics[width=0.43\textwidth]{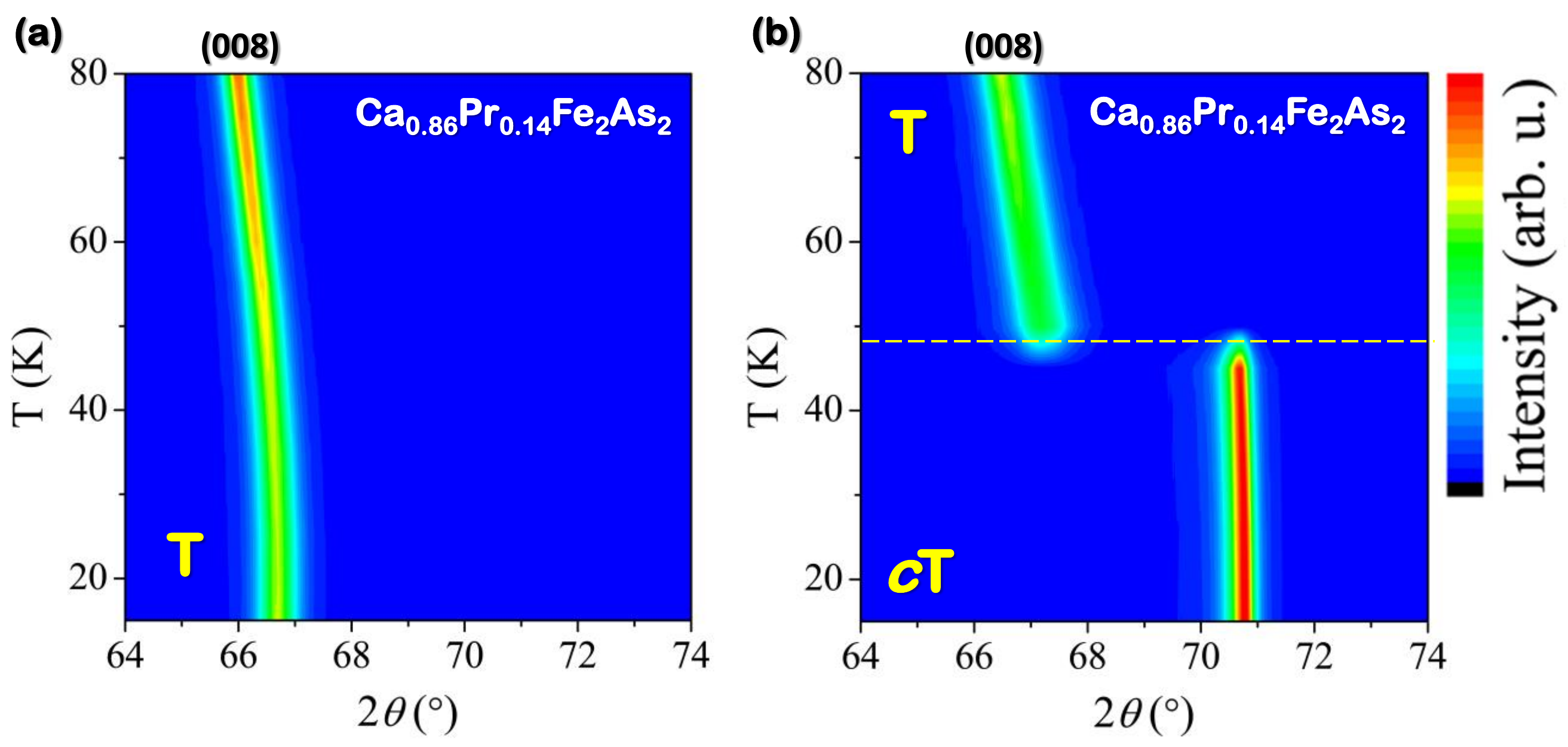}
\caption{(Color online) The temperature dependence of the (008) diffraction peak of Ca$_{0.86}$Pr$_{0.14}$Fe$_{2}$As$_{2}$ for samples that (a) do not show and (b) show the collapsed tetragonal phase transition. The horizontal dashed line marks the transformation region (see text) (see text).}\label{1}
\end{centering}
\end{figure}

\emph{Bulk properties.} -- It has been reported that Ca$_{1-x}$Pr$_{x}$Fe$_{2}$As$_{2}$ superconductor exhibits a transformation from tetragonal ($T$) to non-magnetic collapsed tetragonal ($cT$) phase at $\sim$50-70~K \cite{lv,saha,ma}. We perform x-ray diffraction measurements and use the (008) Bragg peak, collected off the Ca$_{0.86}$Pr$_{0.14}$Fe$_{2}$As$_{2}$ single crystals at high-angle regions, as a probe of the $c$-axis lattice parameter. Surprisingly, the majority ($\sim$80~\%) of our Ca$_{0.86}$Pr$_{0.14}$Fe$_{2}$As$_{2}$ samples do not show the transition (see Fig.\ref{1}), and only a small fraction of our single crystals ($\sim$20~\%) exhibits $T-cT$ transformation at the low temperatures. Notwithstanding the differences, the high $T_{c}$ superconducting state in Ca$_{0.86}$Pr$_{0.14}$Fe$_{2}$As$_{2}$ shows similar characteristics for the samples with or without collapsed tetragonal phase transition (see Supplemental Material in more detail \cite{sm}). Below we focus on the crystals that do not show $T\rightarrow cT$ transition, as they best represent the crystal product of the reaction batch.

\begin{figure}[t!]
\begin{centering}
\includegraphics[width=0.45\textwidth]{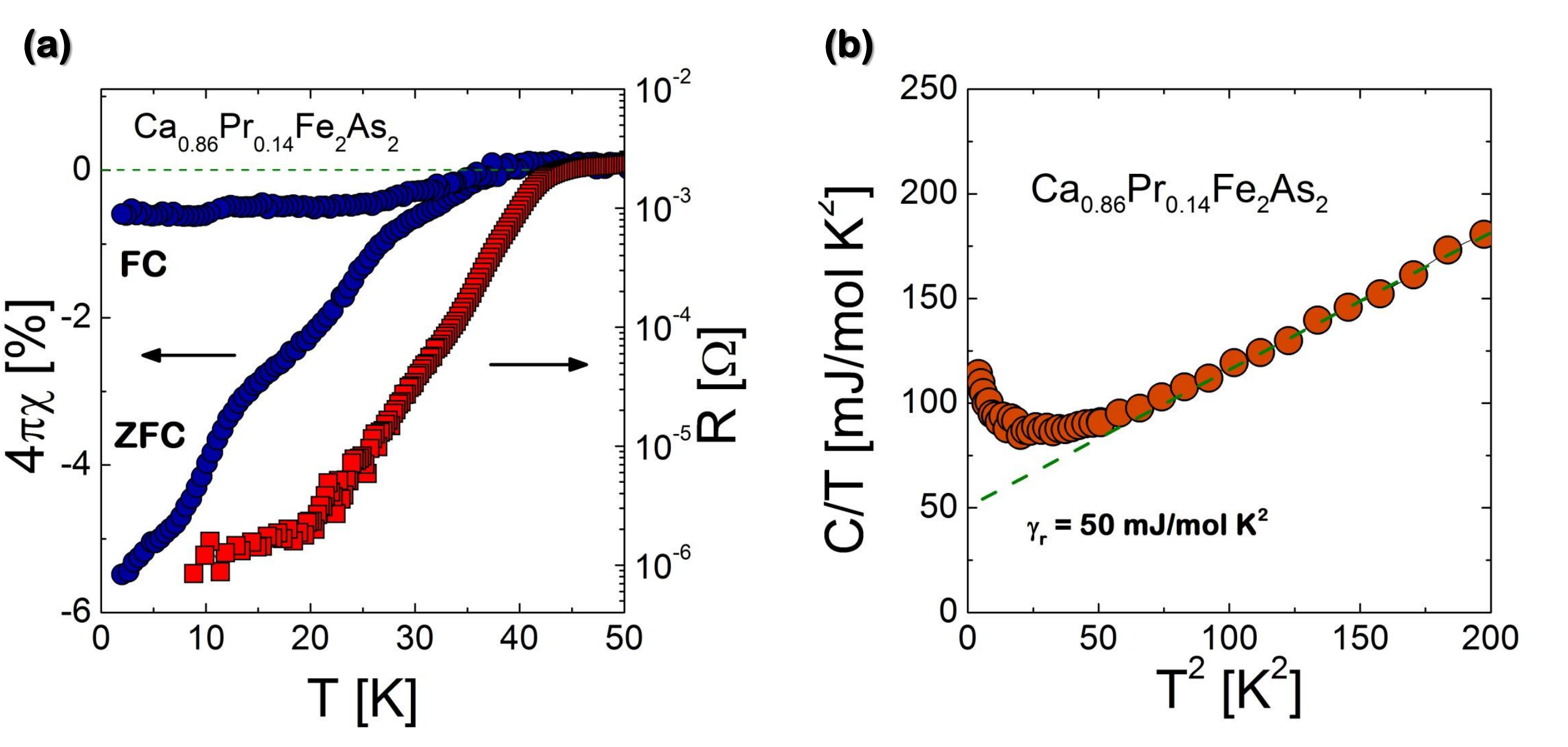}
\caption{(Color online) (a) The normalized magnetic susceptibility measured in zero-field-cooled (ZFC) and field cooled (FC) regimes and the temperature dependence of the electrical resistivity of Ca$_{0.86}$Pr$_{0.14}$Fe$_{2}$As$_{2}$ (b) low temperature specific heat of Ca$_{0.86}$Pr$_{0.14}$Fe$_{2}$As$_{2}$ presented as $C/T$ vs. $T^{2}$ (see text).}\label{2}
\end{centering}
\end{figure}

Typical low-temperature DC magnetic susceptibility and electrical resistivity of Ca$_{0.86}$Pr$_{0.14}$Fe$_{2}$As$_{2}$ single crystals are shown in Fig.\ref{2}a. A diamagnetic behavior, characteristic of superconducting state, is observed in the susceptibility below 45~K together with an additional field repulsion below $\sim$20~K. However, similarly to previous reports, the magnitude of the susceptibility is very small and a maximum superconducting shielding corresponds to $\sim$5\% at 2~K \cite{lv,saha}. The temperature variation of the electrical resistivity of Ca$_{0.86}$Pr$_{0.14}$Fe$_{2}$As$_{2}$ crystals is presented in Fig.\ref{2}a. As seen, the resistivity shows a rapid drop at $T_{c}$~=~45~K (onset $T_{c}$), characteristic of superconducting state. However, the superconducting transition is broad and zero resistance may not even be observed (see Supplemental Material in more detail \cite{sm}). Figure \ref{2}b shows the low temperature dependence the specific heat presented as $C/T$ vs. $T^{2}$. An upturn seen at low temperature is similar to that observed in other Fe-based superconductors and its origin is unclear \cite{greg2,kg}. In addition, the Sommerfeld coefficient which is a measure of the density of states at the Fermi level, is as large as 50~mJ/mol~K$^{2}$. In general, the origin of the residual $\gamma_{r}$ observed in superconducting materials could be caused by pair breaking effects in an unconventional superconductor, crystallographic defects and disorder, and/or spin glass behavior. However, as we show below, a relatively large value of the linear term in the specific heat of Ca$_{0.86}$Pr$_{0.14}$Fe$_{2}$As$_{2}$ is most probably associated with regions of the sample which are not superconducting.

\begin{figure}[t!]
\begin{centering}
\includegraphics[width=0.46\textwidth]{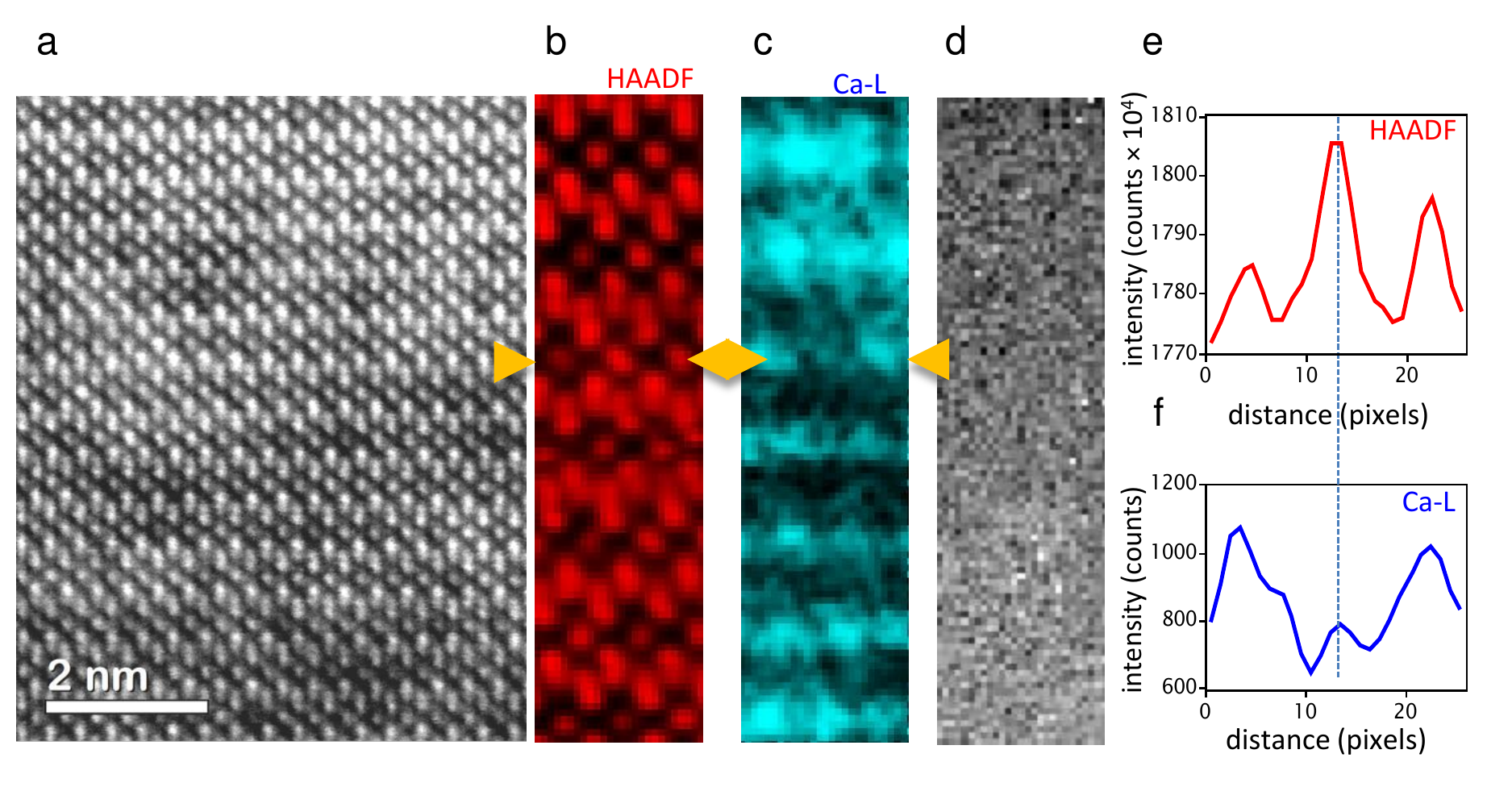}
\includegraphics[width=0.46\textwidth]{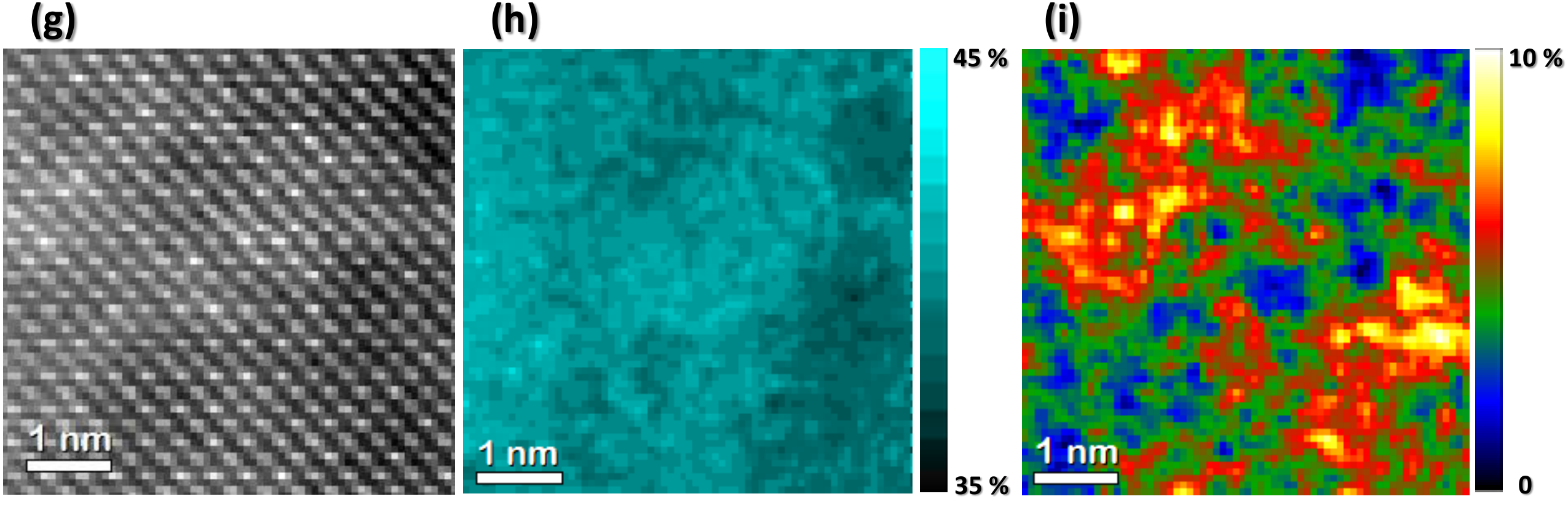}
\caption{(Color online) Sensitivity of STEM/EELS to the Pr distribution: (a) image of Ca$_{0.86}$Pr$_{0.14}$Fe$_{2}$As$_{2}$ along the [100] zone axis. Due to the difference in Z, the atomic columns in the Ca(Pr) plane show different brightness depending on the number of Pr atoms they contain. (b) simultaneous HAADF signal obtained while a spectrum image is acquired by scanning the electron beam over a small region on a 27$\times$100 grid. (c) Ca elemental map obtained by plotting the integrated intensity under the Ca-L$_{2,3}$ edge. (d) thickness map from the zero-loss spectrum for the same region in (b) and (c). (e, f) line scans of (b) and (c) at the position indicated by the arrows. Praseodymium distribution probed by STEM/EELS: (g) simultaneous HAADF image produced while the electron beam scans a region of 13 u.c.$\times$13 u.c. within the \textit{ab} plane using a 70 $\times$ 70 grid. Maps of the Fe (h) and Pr (i) relative concentrations estimated using the Fe-M$_{2,3}$, Ca-L$_{2,3}$, and Pr-N$_{4,5}$ edges (see text).}\label{4}
\end{centering}
\end{figure}

\emph{Nano-scale properties.} -- All the bulk results strongly suggest that nanoscle inhomogeneity may be an important factor in Ca$_{0.86}$Pr$_{0.14}$Fe$_{2}$As$_{2}$. To explore this hypothesis and the origin of the inhomogeneous superconducting state in more detail, we have performed microscopic studies using STEM and STM measurements. Aberration-corrected STEM and EELS provide an adequate tool for probing the distribution of elements within the bulk with angstrom spatial resolution. The best geometry for analyzing the Pr distribution in Ca$_{0.86}$Pr$_{0.14}$Fe$_{2}$As$_{2}$ is a cross sectional one, in which the electron beam is parallel to the [100] axis of the crystal. In this projection, each atomic column contains only one type of atoms: Fe, As, or Ca/Pr. Ca and Pr have significantly different atomic numbers, producing different HAADF intensity. Therefore, the patchy contrast in the HAADF image of Fig.\ref{4}a originates to a large extent from a variation in the number of the Pr atoms present in each Ca/Pr column, and provides a first indication for a non-uniform Pr distribution, which causes dark (Pr-poor) and bright (Pr-rich) patches. Figures~\ref{4}b-f validate this hypothesis, and demonstrate the atomic resolution of EELS in this particular case. Fig~\ref{4}b is another HAADF image in which the beam is now slowly scanned over a coarse grid, allowing for the acquisition of an EEL spectrum for each image pixel (spectrum image). Fig.\ref{4}f shows the Ca map obtained by plotting the Ca-L integrated intensity. We notice that brighter spots in the Ca map correspond to dimmer spots in the simultaneously acquired HAADF image, and therefore to columns with high deficiency of Pr, and vice versa. This is better shown in the line profiles of Fig.\ref{4}e and f, taken at the position indicated by the arrows in Fig.\ref{4}b and c. The small Ca peak in Fig.\ref{4}f correspond to a Ca concentration $\sim$60\% of the Ca concentration in the brightest spots of Fig.\ref{4}c. These variations cannot be attributed to variations in sample thickness, as shown by the thickness map in Fig.\ref{4}d, acquired by using the log-ratio method for the corresponding low-loss EEL spectrum image \cite{RFE}. The thickness of the sample is in fact uniform along each atomic plane in Fig.\ref{4}b, with an average thickness of 0.3 electronic mean free paths. Once established atomic-scale spatial sensitivity of EELS to variations in Pr concentration, more focused spectrum images were acquired as shown in Fig.\ref{4}g. In this case, the spectrum image covers a 13 u.c. $\times$ 13 u.c. region within the basal plane, the same geometry used in STM experiments. The relative Pr concentration is shown in Fig.\ref{4}i, which clearly displays a clustering of Pr atoms within regions of a few u.c. in size, in contrast with the quite uniform distribution for the relative Fe concentration (see Fig.\ref{4}h). We note that the Pr and Fe signals in the plane view maps do not show atomic resolution. This is due in part to the coarser grid used and beam broadening effects, but also to the larger disorder associated with the (001) surface, which is observed also in STM and explains the small ($\sim$3\%) variation in the Fe relative concentration. Despite the signal delocalization, a simple statistic of Fig.\ref{4}i reveals that, if we assume the Pr atoms are located only in the regions with intensity above the mean value (colored red to white in Fig.\ref{4}i), and ignore contributions from low-intensity pixels below the 25$^{th}$ percentile, the biggest cluster in Fig.\ref{4}i would contain $\sim$10 Pr atoms. Considering that the electron beam probes a thickness of about 8 unit cells along $c$, a monolayer slice of the sample in the $ab$ plane would show very small Pr nuclei containing only a few atoms and spaced only a few unit cells apart.

\begin{figure}[t!]
\begin{centering}
\includegraphics[width=0.45\textwidth]{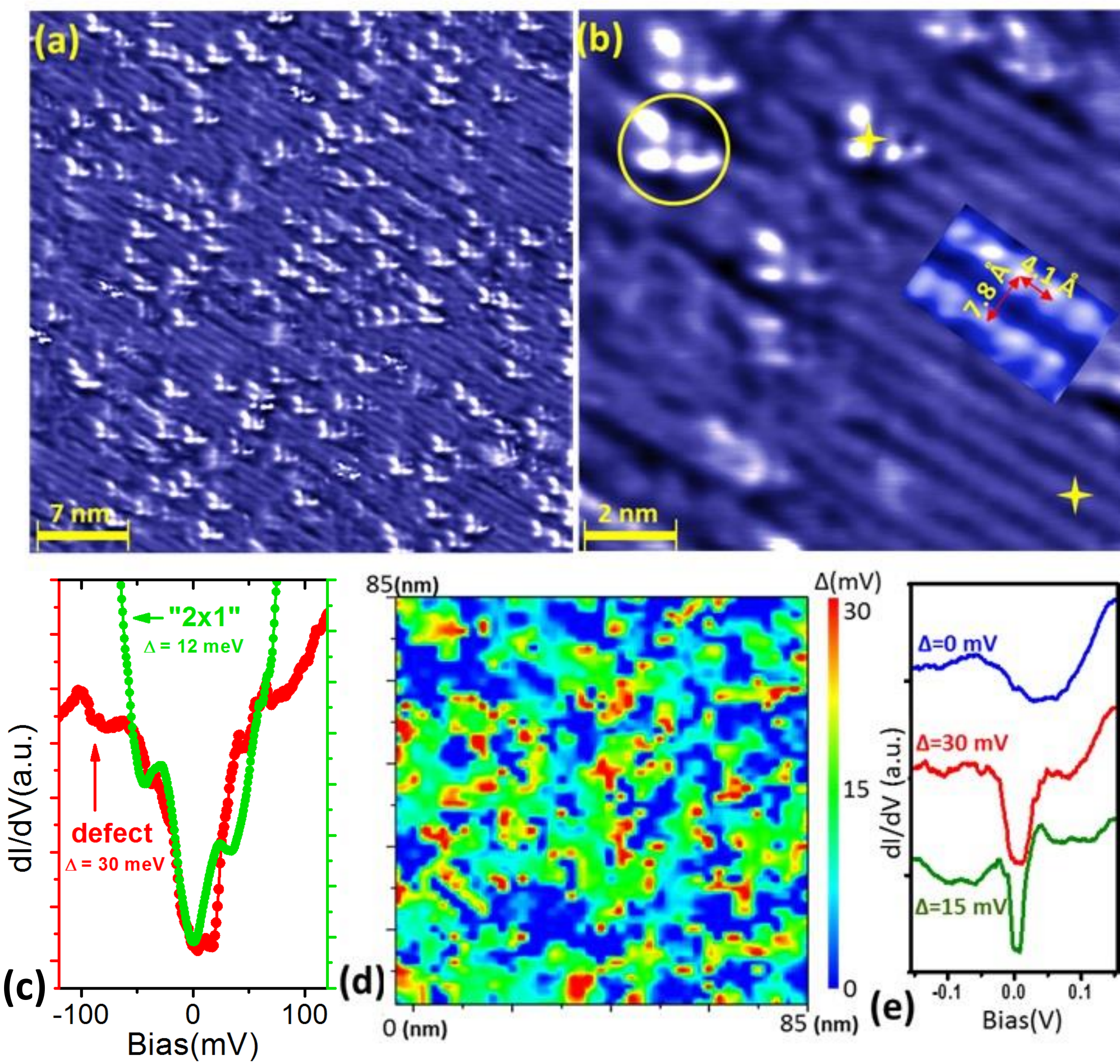}
\caption{(Color online) Surface morphologies of in situ-cleaved Ca$_{0.86}$Pr$_{0.14}$Fe$_{2}$As$_{2}$ single crystal and corresponding superconducting gaps: (a) 42~nm$\times$42~nm STM topographic image of the cleaved surface, taken with bias V = -0.3~V and $I_{t}$~=~50~pA. The STM image shown here is flattened to remove the overall roughness and enhance the atomic contrast of dopants. (b) Higher resolution STM image showing local 2$\times$1 structure. (c) $dI/dV$ curves measured on clover-like defect (red) and on ordered 2$\times$1 region (green). The locations are marked by yellow stars in panel b. (d) Superconducting gap map derived from $dI/dV$ spectra grid over a larger area of 84~nm$\times$84~nm size. (e) Three representative $dI/dV$ spectra for non-superconducting (blue) and superconducting (green, red) regions. The $dI/dV$ spectra grid was taken with V = -100~mV, $I_{t}$~= 10~pA, and modulation 3~mV (see text).}\label{6}
\end{centering}
\end{figure}

Figure~\ref{6}a displays a typical topographic image taken at 4.2~K for the in situ-cleaved Ca$_{0.86}$Pr$_{0.14}$Fe$_{2}$As$_{2}$ single crystal. To visualize the defects, we flatten the STM image by compressing the overall roughness of these bright spots in order to enhance the atomic contrast of these dopants. As shown in this large scale image (42~nm$\times$42~nm), the majority of the sample surface is covered by the so-called 2$\times$1 stripes. As the in-plane lattice constant is 3.98~{\AA}, a high resolution image of two stripes, inset of Fig.\ref{6}b, reveals a 2$\times$1 (7.8~{\AA}$\times$4.1~{\AA}) structure. Such 2$\times$1 striped network has been reported frequently on the surface of various 122 Fe-based superconductors, resulting from a half layer of Ba/Sr/Ca after cleaving (see Refs.\onlinecite{1,2,3,4,5}). Furthermore, clover-like defects, around 1-2~nm in size, are found disrupting the 2$\times$1 striped structure. Figure~\ref{6}b shows that each defects consist of four atomic features attached with three bright tails. Under different bias, the defects turn into bright spots (see Supplemental Material in more detail \cite{sm}), similar to the observations of Zeljkovic \textit{et al.}~\cite{6}, where such defects are identified to be Pr-dopants. Interestingly, a similar clover-like structure has been recently observed in doped Bi$_{2}$Se$_{3}$ topological insulator \cite{STM1,STM2}. In the surrounding area of these defects, the surface lattice becomes highly disordered in comparison to the unperturbed 2$\times$1 stripe structure. Such "disordered" region merges smoothly into the 2$\times$1 structure, and is likely a reconstruction of the Ca layer \cite{6}.

Point tunneling spectra taken at 4.2~K on ordered 2$\times$1 regions (see lower star in Fig.\ref{6}b) demonstrates a clear superconducting gap (green curve in Fig.\ref{6}c). The STS curves can be fit to the Dynes function \cite{8} using a gap value of $\Delta\sim$12~meV, yielding the ratio 2$\Delta$/k$_{B}T_{c}$ $\sim$~6.3. This is consistent with reported values for other 122 Fe-based superconductors \cite{9,10} and it is also expected for a moderate-coupling BCS superconductor. However, such a gap structure varies spatially; the STS curves measured on the clover-like defect (see upper star in Fig.\ref{6}b) show a gap size as large as 30~meV with no prominent coherence peaks (red curve in Fig.\ref{6}c). To further investigate the gap distribution we took a differential tunneling conductance spectra ($dI/dV$ versus $V$) survey on a large surface area, using the previously explored method \cite{11,12,13} (see Fig.\ref{6}d). Fig.\ref{6}e shows three representative tunneling spectra for superconducting (red and green curves) and normal state (blue curve) regions. The gap map demonstrates a distinct nanoscale phase separation between superconducting and non-superconducting regions. Over 2/3 area (green color) has a gap around 15~meV, while $\sim$1/3 of the material is in normal state at 4.2~K. The superconducting regions are connected to each other, therefore the presence of the percolating superconducting path and zero resistance is most probably determined by the green regions in Fig.\ref{6}d. Interestingly, a localized regions of strong superconductivity are observed with a gap as large as 30~meV (see red areas in Fig.\ref{6}d). The $\Delta$~=~30~meV gap of the red regions in Fig.\ref{6}d matches the gap value obtained on the clover-like defect (see red curve in Fig.\ref{6}c). Therefore, these defects that are associated with praseodymium dopants create regions of strong localized superconductivity in this material. Gap inhomogeneity exists in most doped superconductors, including Fe-based materials, however, the majority of the materials are bulk superconductors, showing only nanoscale variation of the gap (see Refs.~\onlinecite{CUP1,CUP2,co1,13,6,co3,co4}). The appearance of a non-superconductive phase and related phase separation explains the filamentary nature of the superconductivity in this material and is in agreement with the bulk measurements. The STM and STEM/EELS data, together with previous studies on this material \cite{6} indicate that the clover-like defects observed in the topographic image can be identified with Pr dopants, and mark the site of localized regions of strong high-$T_{c}$ superconductivity. Furthermore, the size of the large-gap red regions in Fig.\ref{6}d is very similar to the size of the red/yellow clusters in Fig.\ref{4}i, indicating that the Pr clusters are indeed very small, composed of 3 or 4 atoms, and separated from other clusters by a few unit cells.

\emph{Summary and outlook.} -- In summary, we have investigated the effects of electronic inhomogeneity and filamentary superconductivity in Ca$_{0.86}$Pr$_{0.14}$Fe$_{2}$As$_{2}$. We use an extensive macro- and micro-probe measurements and show that the inhomogeneity is manifested as a spacial variation of both, local density of states, and the superconducting order parameter. Our analysis shows that the inhomogeneous and strongly localized high-$T_{c}$ superconducting state emerges from clover-like defects, and is a consequence of a Pr distribution which is non uniform although on a very small scale. In addition, a significant part of the sample remains in normal state at 4.2~K, in agreement with the bulk studies. Interestingly, the presence of the high-$T_{c}$ superconductivity in Ca$_{0.86}$Pr$_{0.14}$Fe$_{2}$As$_{2}$ is observed in both, tetragonal and collapsed tetragonal phases. It has been shown recently that the electronic structure is strongly reconstructed below the $cT$ transition in CaFe$_{2}$As$_{2}$, leading to disappearance of the hole pocket and magnetism at the center of the Brillouin zone \cite{kg2,ty}. If similar situation is observed in Pr-doped CaFe$_{2}$As$_{2}$, this would warrant a question on the nature of superconductivity and the role of interband spin fluctuation pairing mechanism in this material, and in other unconventional superconductors.

\begin{acknowledgments}

This work was supported by the Department of Energy, Basic Energy Sciences, Materials Sciences and Engineering Division.

\end{acknowledgments}

\end{document}




\title{Supplemental Material for \emph{"Local inhomogeneity and filamentary superconductivity in Pr-doped CaFe$_{2}$As$_{2}$"}}

\author{K. Gofryk}
\author{M.~Pan}
\author{C.~Cantoni}
\author{B.~Saparov}
\author{J.~E.~Mitchell}
\author{A.~S.~Sefat}
\affiliation{Oak Ridge National Laboratory, Oak Ridge, Tennessee 37831, USA}

\date{\today}

\maketitle

\textbf{In this Supplement we provide additional information regarding: (i) transport properties of Ca$_{0.86}$Pr$_{0.14}$Fe$_{2}$As$_{2}$ samples with and without collapsed tetragonal phase transition  and (ii) STM measurements of Ca$_{0.86}$Pr$_{0.14}$Fe$_{2}$As$_{2}$.}\\

\section{Transport properties}

As stated in the manuscript, the majority ($\sim$80~\%) of our Ca$_{0.86}$Pr$_{0.14}$Fe$_{2}$As$_{2}$ samples did not show the collapsed tetragonal phase transition, and only a small fraction ($\sim$20~\%) exhibits $T-cT$ transformation at low temperatures. Figs.S1a and b show the temperature dependence of the (008) Bragg peak, collected off an Ca$_{0.86}$Pr$_{0.14}$Fe$_{2}$As$_{2}$ single crystals at high-angle regions, as a probe of $c$-axis lattice parameter. For samples with the $cT$ transition (Fig.S1a), the (008) peak increases with lowering temperature, characteristic of formation of the collapsed tetragonal phase at about 50~K. The absence of this transformation in other Ca$_{0.86}$Pr$_{0.14}$Fe$_{2}$As$_{2}$ single crystal can be seen in Fig.S1b. The Figs.S1c and d show the temperature dependence of the electrical resistivity measured for these two particular samples. In both pieces of crystal, a drop of the resistivity is observed at $\sim$45~K (onset $T_{c}$) characteristic of superconducting state. As can be seen from the figures an additional transition is observed at $\sim$20~K and for some samples a zero resistance value has been observed below 20~K, as exemplified in Fig.S1d. The temperature dependence of the electrical resistivity, measured for several values of the magnetic field applied parallel to the \textit{c}-axis of the crystal of Ca$_{0.86}$Pr$_{0.14}$Fe$_{2}$As$_{2}$ is shown in figures S1e and f. The superconducting anomaly is gradually depressed and the transition is broadened with the field applied. As seen from the figures, the magnetic field has a slightly different effect on the superconducting state to the samples with and without the collapsed tetragonal phase transition. Figs.S1g and h show the temperature dependence of derivative of the electrical resistivity of those two samples at different magnetic fields. The $d\rho/dT$ curves show broad anomalies consistent with the presence of the inhomogeneous superconducting state and significant spacial distribution of $T_{c}$ in Ca$_{0.86}$Pr$_{0.14}$Fe$_{2}$As$_{2}$ (see the main text). We track the most prominent anomalies (see arrows in Figs.S1g and h) and plot $H$ vs. $T$ in Figs.S1i and j. As seen, for both samples $T_{c}$ (defined at a maximum in $d\rho/dT$) decreases upon applied magnetic field with the slope of $\frac{dH_{c2}}{dT}$~=~-6~T~K$^{-1}$. From the slope of the critical field, at high magnetic fields, the value of $H_{c2}$ may be evaluated by the Werthammer-Helfand-Hohenberg (WHH) model.\cite{WHH1,WHH2} In this approach (taking into account the orbital effect only; Maki parameter $\alpha$~=~0) the critical field may be expressed as $H_{c2}^{orb}$~=~0.69$\times~T_{c}\left[\frac{dH_{c2}}{dT}\right]_{T_{c}}$ and for Ca$_{0.86}$Pr$_{0.14}$Fe$_{2}$As$_{2}$ this gives ~180~T. Interestingly, for sample showing the collapsed tetragonal transition another anomaly having $\frac{dH_{c2}}{dT}$~=~-2~T~K$^{-1}$ and $H_{c2}$~=~59~T can be identified and tracked.

All the results obtained indicate the presence of the filamentary and inhomogeneous superconducting state in Ca$_{0.86}$Pr$_{0.14}$Fe$_{2}$As$_{2}$ with significant spacial distribution of $T_{c}$. Furthermore, the transport data might suggest larger amount of the high temperature superconducting phase in in Ca$_{0.86}$Pr$_{0.14}$Fe$_{2}$As$_{2}$ is present in the tetragonal crystal structure, as compared to the collapsed tetragonal one. At this point we don't know the origin of this behaviour. This could be simply associated with a different degree of inhomogeneity in these two samples, however it could also have a more fundamental origin, related to the nature of the superconductivity in Pr-doped CaFe$_{2}$As$_{2}$ (see the main text).

\section{STM measurements}

Figure S2 shows a large-scale topographic STM image acquired at 4~K for the in situ-cleaved Pr-doped CaFe$_{2}$As$_{2}$ single crystal.
The size is 84 nm $\times$ 84 nm, taken with bias = -0.5 V. In addition to linear stripe structure existing all over the surface, bright spots with similar size and uniform distribution were also observed, similar to the observations of Zeljkovic \textit{et al.} \cite{6}, where such defects are believed as Pr-dopant. As shown in the paper, the the bright spots actually have a clover-like structure in fllatten STM image.

\begin{figure*}
\begin{minipage}{1\textwidth}
\includegraphics[width=0.6\textwidth]{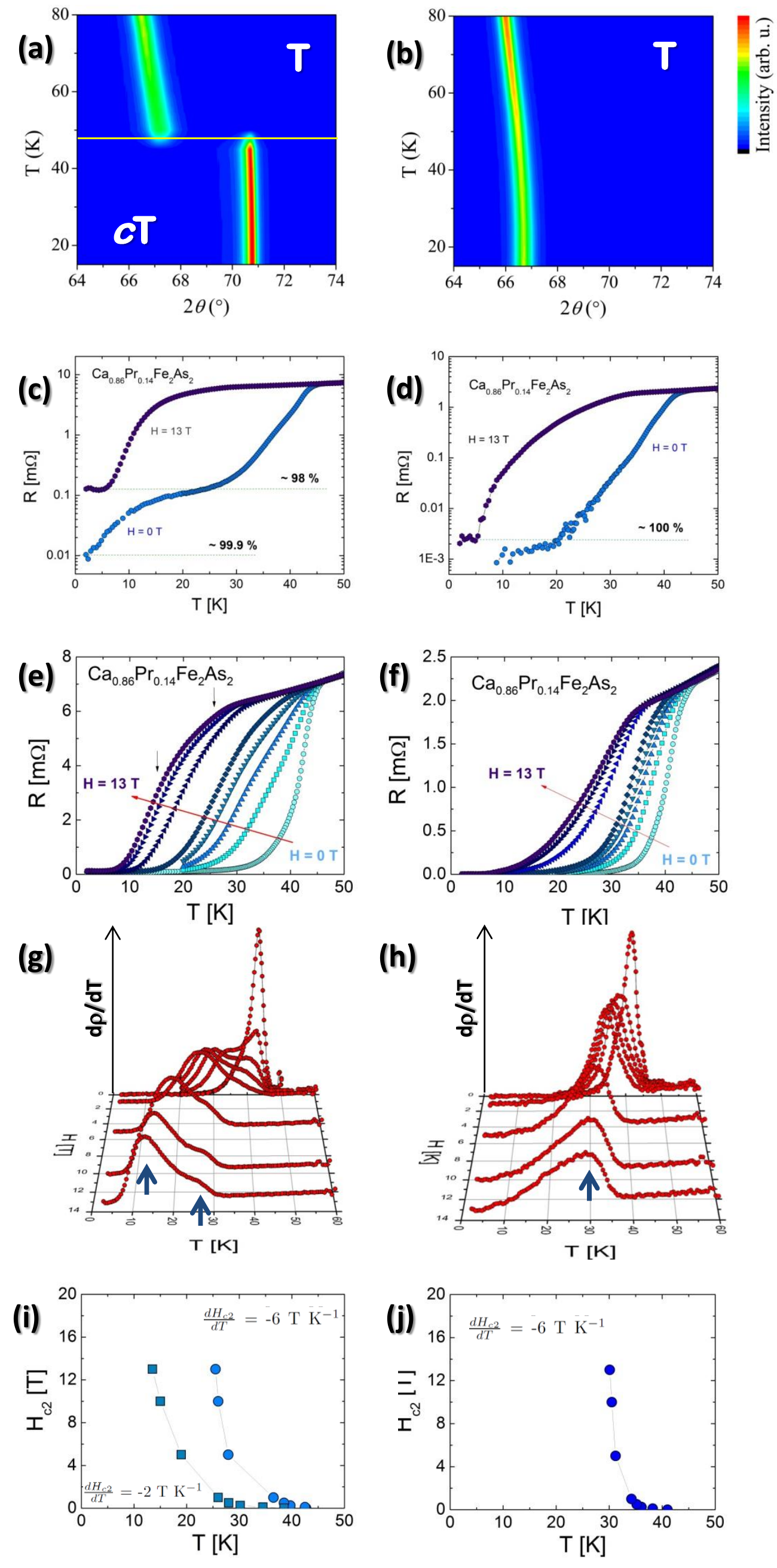}
\end{minipage}
\setlength{\unitlength}{1in}
\begin{picture}(0,0)(0,0)
\end{picture} \vspace{0.17cm}
\caption{(Color online) The crystallographic and transport data measured on the same two pieces of Ca$_{0.86}$Pr$_{0.14}$Fe$_{2}$As$_{2}$ single crystals. The temperature dependence of the (008) diffraction peak of Ca$_{0.86}$Pr$_{0.14}$Fe$_{2}$As$_{2}$ illustrates for (a) transformation from high temperature tetragonal phase ($T$) to low temperature collapsed tetragonal phase ($cT$) and for (b) lack of this transformation for this sample. The horizontal dashed lines mark the transformation region. (c, d) The temperature dependence of the electrical resistivity for these two samples. (e, f) $\rho(T)$ measured at several different magnetic fields. (g, h) The temperature and magnetic field dependencies of the derivative of the electrical resistivity. Arrows mark the anomalies at $d\rho/dT$. (i, j) $H_{c2}$ vs. $T$ (see text).}\label{2}
\end{figure*}

\begin{figure*}
\begin{minipage}{1\textwidth}
\includegraphics[width=0.8\textwidth]{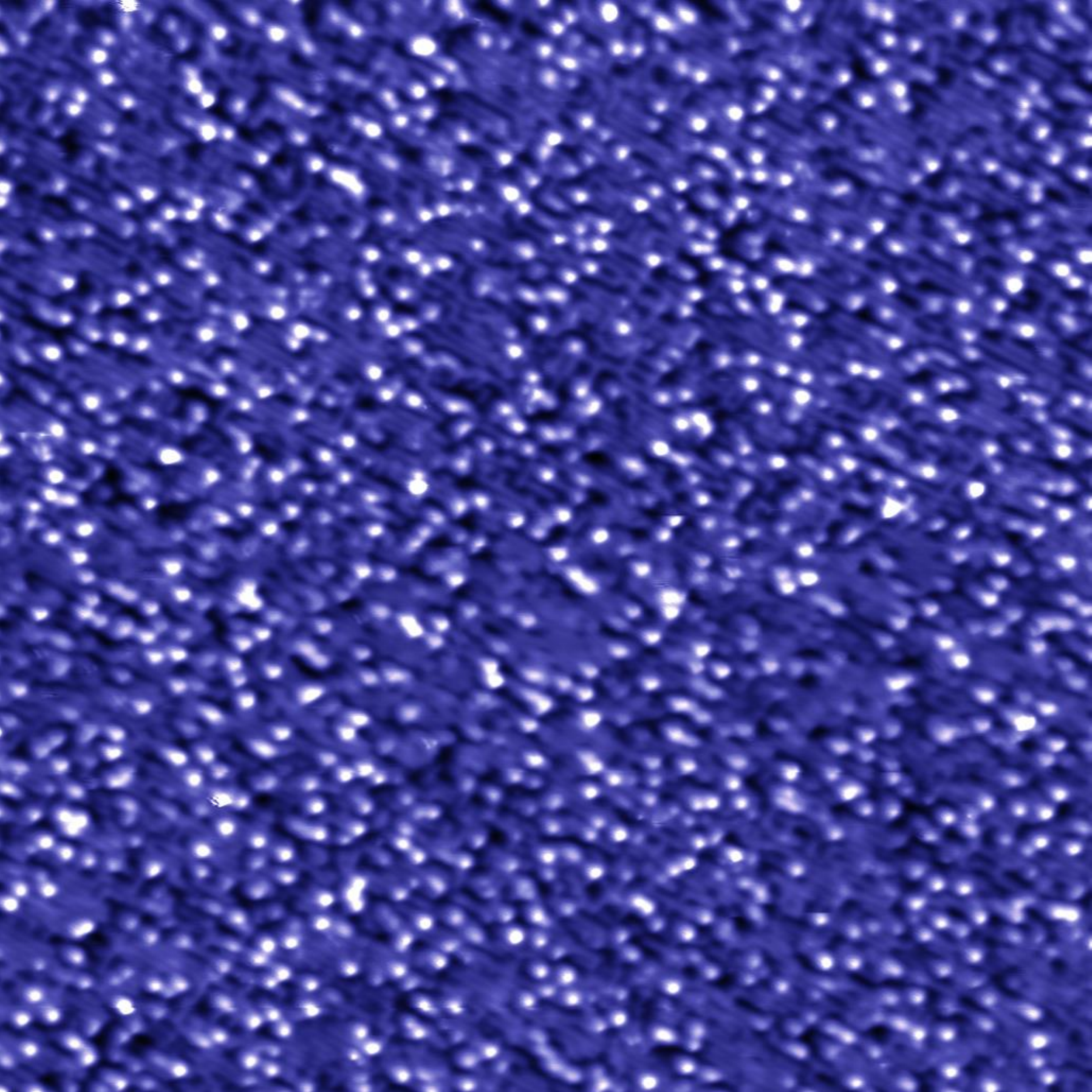}
\end{minipage}
\setlength{\unitlength}{1in}
\begin{picture}(0,0)(0,0)
\end{picture} \vspace{0.17cm}
\caption{(Color online) A large scale, topographic STM image (raw) taken at 4~K for the in situ-cleaved Ca$_{0.86}$Pr$_{0.14}$Fe$_{2}$As$_{2}$ single crystal sample. The size is 84 nm $\times$ 84 nm, taken with bias = -0.5 V (see text).}\label{1}
\end{figure*}